\documentclass[amsmath,amssymb,aps,prc,preprint,superscriptaddress]{revtex4-1}

\usepackage{color}
\usepackage{graphicx}

\begin{document}


\title{
Photodisintegration cross section of $^9$Be up to 16 MeV in the $\alpha$~+~$\alpha$~+~$n$ three-body model
}


\author{Yuma Kikuchi}
\affiliation{RIKEN Nishina Center, Wako 351-0198, Japan}
\email[]{yuma.kikuchi@riken.jp}
\author{Myagmarjav Odsuren}
\affiliation{School of Engineering and Applied Sciences, National University of Mongolia, Ulaanbaatar 14200, Mongolia}
\author{Takayuki Myo}
\affiliation{General Education, Faculty of Engineering, Osaka Institute of Technology, Osaka 535-8585, Japan}
\affiliation{Research Center for Nuclear Physics (RCNP), Osaka University, Ibaraki 567-0047, Japan}
\author{Kiyoshi Kat\=o}
\affiliation{Nuclear Reaction Data Centre, Faculty of Science, Hokkaido University, Sapporo 060-0810, Japan}



\date{\today}

\begin{abstract}
The photodisintegration of $^9$Be in the energy region lower than $E_\gamma = 16$ MeV is investigated by using the $\alpha$~+~$\alpha$~+~$n$ three-body model and the complex scaling method.
The cross section exhibits two aspects in the different two energy regions.
In the low energy region up to $E_\gamma = 6$ MeV, the cross section is explained by the transition strengths into the excited resonant states of $^9$Be, while the dipole transition into the non-resonant continuum states of $^8$Be(2$^+$)~+~$n$ dominates the cross section in the energy region of $6 \le E_\gamma \le 16$ MeV.
Furthermore, it is shown that the dipole strength at $E_\gamma \sim 8$ MeV is understood to be caused by the single-neutron excitation from the $^8$Be(2$^+$)~$\otimes$~$\nu p_{3/2}$ configuration in the ground state. 
\end{abstract}


\maketitle

\section{Introduction}
The photodisintegration cross section of $^9$Be shows various characters in different energy regions.
The cross section in a low energy region has been measured to deduce a production rate of $^9$Be from the astrophysical point of view~\cite{Gibbons59,John62,Fujishiro82,Utsunomiya00,Burda10,Arnold12,Utsunomiya15}.
In the low energy region up to $E_\gamma = 6$ MeV, the cross section has been observed to come from the electromagnetic transitions into the excited states of $^9$Be, and theoretically, has been studied within the $\alpha$~+~$\alpha$~+~$n$ three-body model~\cite{Garrido10,A_Rodriguez10,Casal14,Odsuren15}.
In particular, the first excited $1/2^+$ state, observed as a sharp peak just above the  $^8$Be($0^+$)~+~$n$ threshold, has attracted much attention, and its structure has been discussed in terms of the three-body resonance of $\alpha$~+~$\alpha$~+~$n$~\cite{Garrido10,A_Rodriguez10,Efros14} or the two-body virtual state of $^8$Be~+~$n$~\cite{Efros99,Arai03,Odsuren15}.

In the energy region higher than $E_\gamma = 6$ MeV, the recent measurement~\cite{Utsunomiya15} reports the photodisintegration cross section of $^9$Be up tp $E_\gamma = 16$ MeV, and the cross section shows a significant electric dipole strength below the giant dipole resonance (GDR).
An enhancement of a low-lying dipole strength below the GDR has been observed in a wide range of the mass number in neutron-rich nuclei and is often denoted by the pygmy dipole resonance (PDR).
The PDR has been discussed in neutron-rich nuclei in relation with the neutron skin thickness.
On the other hand, $^9$Be is a light and stable nucleus which has been studied to have a cluster structure, and hence, it is interesting to investigate the mechanism of the dipole strength in $^9$Be below the GDR energy region.
In the recent measurement~\cite{Utsunomiya15}, Utsunomiya {\it et al.} discussed that the enhanced dipole strength in $^9$Be at the excitation energy of $\sim$ 8 MeV exhausts 10~\% of Thomas-Reiche-Kuhn sum rule and almost all the cluster dipole sum rule~\cite{Utsunomiya15}.
It is desired to understand the low-lying dipole strength below the GDR comprehensively from a viewpoint of the cluster structure of $\alpha$~+~$\alpha$~+~$n$ in $^9$Be.

In our previous work, we have investigated the structure of the $1/2^+$ state of $^9$Be located just above the $\alpha$~+~$\alpha$~+~$n$ threshold energy using the $\alpha$~+~$\alpha$~+~$n$ three-body model and the complex scaling method (CSM)~\cite{Odsuren15}.
We calculated the photodisintegration cross section from the ground state into the $1/2^+$ state and reproduce the observed peak in the cross section just above the $^8$Be($0^+$)~+~$n$ threshold.
On the other hand, we could not find a sharp resonance corresponding to the $1/2^+$ peak in the cross section using the CSM calculation.
To understand the origin of the $1/2^+$ peak in the cross section, we performed the calculation of the pole trajectory by changing the attraction of the inter-cluster force as an analytical continuation.
From these analyses, we concluded that the excited $1/2^+$ state has a virtual-state character of the $s$-wave neutron in the $^8$Be~+~$n$ system.

In Ref.~\cite{Odsuren15}, we focused our discussion on the structure of the $1/2^+$ state and its contribution to the photodisintegration cross section in the low energy region.
Other spin-parity states than the $1/2^+$ state were not included in the calculation and the photodisintegration cross section in higher energy regions was not discussed.
In order to understand the feature of $^9$Be, it is necessary to examine the photodisintegration cross section by including all the available spin-parity states connected with the ground $3/2^-$ state via electromagnetic transitions and by taking into account transitions in higher energy regions.

The purposes of this work are following two:
One is to investigate the excited states of $^9$Be in low energy region connected with the ground state through the electromagnetic transitions.
The other is to elucidate the mechanism of the enhanced dipole transition in $^9$Be below the GDR observed by the recent experiment~\cite{Utsunomiya15}. 
In the present calculation, we employ the $\alpha$~+~$\alpha$~+~$n$ three-body model with complex-range Gaussian basis functions to describe the scattering states of the $\alpha$~+~$\alpha$~+$n$ system retaining the numerical accuracy.
We calculate the photodisintegration cross section applying the CSM with Green's function to the $\alpha$~+~$\alpha$~+~$n$ three-body model, and discuss the mechanism of the photodisintegration cross section .

This article is organized as follows:
In Sec.~\ref{sec:f}, we explain the $\alpha$~+~$\alpha$~+~$n$ three-body model and describe the formalism of the photodisintegration cross section using the CSM.
In Sec.~\ref{sec:r}, we show the results of the photodisinegration cross section, and discuss the structure of the excited states and the mechanism of the low-lying dipole strength in $^9$Be up to $E_\gamma = 16$ MeV.
Finally, in Sec.~\ref{sec:s}, all results and discussions are summarized.

\section{Theoretical framework\label{sec:f}}
\subsection{$\alpha$~+~$\alpha$~+~$n$ three-body model for $^9$Be}
We briefly explain the $\alpha$~+~$\alpha$~+~$n$ three-body model employed in the present work, whose detail is given in Ref.~\cite{Odsuren15}.
We here solve the Schr\"odinger equation for the $\alpha$~+~$\alpha$~+~$n$ system using the orthogonality condition model~\cite{Saito77}.
The Schr\"odinger equation is given as
\begin{equation}
\hat{H}\Psi^\nu_{J^\pi}=E_\nu\Psi^\nu_{J^\pi},
\label{eq:scheq}
\end{equation}
where $J^\pi$ is the total spin and parity of the $\alpha$~+~$\alpha$~+~$n$ system and $\nu$ is the state index.
The energy eigenvalue $E_\nu$ is measured from the $\alpha$~+~$\alpha$~+$n$ threshold.

The Hamiltonian for the relative motion of the $\alpha$~+~$\alpha$~+~$n$ three-body system for $^9$Be is given as
\begin{equation}
\hat{H}=\sum^{3}_{i=1}t_{i}-T_\text{c.m.}+\sum^{2}_{i=1}V_{\alpha n}(\boldsymbol{\xi}_{i})+V_{\alpha\alpha}+V_\text{PF}+V_3,
 \label{eq:ham}
\end{equation}
where $t_i$ and $T_\text{c.m.}$ are kinetic operators for each particle and the center-of-mass of the system, respectively.
The interaction between the neutron and the $i$-th $\alpha$ particle is given as $V_{\alpha n}(\boldsymbol{\xi}_i)$, where $\boldsymbol{\xi}_i$ is the relative coordinate between them.
We here employ the KKNN potential~\cite{Kanada79} for $V_{\alpha n}$.
For the $\alpha$-$\alpha$ interaction $V_{\alpha \alpha}$ we employ the same potential as used in Ref.~\cite{Odsuren15}, which is a folding potential of the effective $NN$ interaction~\cite{Schmid61} and the Coulomb interaction.
The pseudo potential $V_\text{PF}= \lambda|\Phi_\text{PF}\rangle\langle\Phi_\text{PF}|$ is the projection operator to remove the Pauli forbidden states from the relative motions of $\alpha$-$\alpha$ and $\alpha$-$n$~\cite{Kukulin86}.
The Pauli forbidden state $\Phi_\text{PF}$ is defined as the harmonic oscillator wave functions by assuming the $(0s)^4$ configuration whose oscillator length is fixed to reproduce the observed charge radius of the $\alpha$ particle.
In the present calculation, $\lambda$ is taken as $10^6$ MeV.

In the present calculation, we introduces the $\alpha$~+~$\alpha$~+~$n$ three-body potential $V_3$. The explicit form of $V_3$ is given as
\begin{equation}
V_3 = v_3 \exp{(- \mu \rho^2)},
\label{eq:3bp}
\end{equation}
where $\rho$ is the hyperradius of the $\alpha$~+~$\alpha$~+~$n$ system.
The hyperradius is defined as
\begin{equation}
\rho^2 = 2r^2 + \frac{8}{9}R^2,
\end{equation} 
where $r$ is the distance between two $\alpha$'s and $R$ is that between the neutron and the center-of-mass of the $\alpha$~+~$\alpha$ subsystem.
The strength and width of the three-body potential, $v_3$ and $\mu$, are determined for each spin and parity.
For $3/2^-$ states, we determine the parameters to reproduce the observed binding energy and charge radii of the ground state because it is essential to reproduce the $Q$-value and the sum rule value of the electric dipole transition in discussing the photodisintegration of $^9$Be.
To reproduce the ground-state properties, we take $v_3$ and $\mu$ as 1.10 MeV and 0.02 fm$^{-2}$, respectively.
For other spin-parity states, we employ the same value of $\mu$ as used in $3/2^-$ states, and different strengths are used to reproduce the energy positions of the observed peaks in the photodisintegration cross section.

We solve the Schr\"odinger equation with the coupled-rearrangement-channel Gaussian expansion method~\cite{hiyama03}.
In the present calculation, the $^9$Be wave function $\Psi^\nu_{J^\pi}$ is described in the Jacobi coordinate system as
\begin{equation}
\Psi^\nu_{J^\pi} = \sum_{ijc} C_{ijc}^\nu(J^\pi) \left[\left[\phi^i_l(\mathbf{r}_c),\phi^j_\lambda(\mathbf{R}_c)\right]_L,\chi_{\frac{1}{2}}\right]_{J^\pi},
\label{eq:basis}
\end{equation}
where $C_{ijc}^\nu(J^\pi)$ is a expansion coefficient and $\chi_{\frac{1}{2}}$ is the spin wave function.
The relative coordinates $\mathbf{r}_c$ and $\mathbf{R}_c$ are those in three kinds of the Jacobi coordinate systems indexed by $c$ $(=1,2,3)$, and the indices for the basis functions are represented as $i$ and $j$.
The spatial part of the wave function is expanded with the complex-range Gaussian basis functions~\cite{hiyama03}.
The explicit forms of the complex-range Gaussian basis functions are given as
\begin{equation}
\phi^i_l (\mathbf{r}) =
\begin{cases}
N^S_l(a_i) \exp\left(-a_i r^2\right) \sin{\left(a_i\omega r^2\right)}\\
N^C_l(a_i) \exp\left(-a_i r^2\right) \cos{\left(a_i\omega r^2\right)},
\end{cases}
\label{eq:CGgauss}
\end{equation}
where $N^S_l$ and $N^C_l$ are the normalization factors and $a_i$ is a width of the Gaussian basis function.
The basis functions in Eq.~(\ref{eq:CGgauss}) enables us to treat the oscillating behavior in the relative motion and is useful to describe the photodisintegration cross section accurately.

\subsection{Photodisintegration cross section in the complex scaling method}

To calculate the photodisintegration cross section, we use the complex scaling method (CSM)~\cite{Aguilar71,Balslev71,Ho83, Moiseyev98, Aoyama06,Myo14,Odsuren15}.
In the CSM, the relative coordinates $\boldsymbol{\xi}$ ($\mathbf{r}_c$ and $\mathbf{R}_c$) are transformed as
\begin{equation}
U(\theta)\boldsymbol{\xi} U^{-1}(\theta) = \boldsymbol{\xi}e^{i\theta},
\end{equation}
where $U(\theta)$ is a complex scaling operator and $\theta$ is a scaling angle being a real number.
Applying this transformation to Eq.~(\ref{eq:scheq}), we obtain the complex-scaled Schr\"odinger equation as
\begin{equation}
\hat{H}^\theta \Psi^\nu_{J^\pi} (\theta) = E^\theta_\nu \Psi^\nu_{J^\pi} (\theta).
\label{eq:CSsch}
\end{equation}
The complex-scaled Hamiltonian $\hat{H}^\theta$ and the complex-scaled wave function $\Psi^\nu_{J^\pi}(\theta)$ are defined as in Ref.~\cite{Myo14}.
By solving the complex-scaled Schr\"odinger equation with the $\mathcal{L}^2$ basis function given in Eq.~(\ref{eq:CGgauss}), we obtain the energy eigenvalues and eigenstates (their biorthogonal states) as $\{E^\theta_\nu\}$ and $\{\Psi^\nu_J(\theta)\}$ ($\{\tilde{\Psi}^\nu_J(\theta)\}$), respectively.

When we apply the CSM to the complex-range Gaussian basis functions, we need to choose the value of the scaling angle $\theta$ carefully in relation with the parameter $\omega$ in Eq.~(\ref{eq:CGgauss}).
Applying the CSM to the complex-range Gaussian basis function, we obtain
\begin{equation}
\begin{split}
&\exp{\left( - a r^2 e^{2i\theta} \right)}\cos{\left( a \omega r^2e^{2i\theta} \right)} \\
&= \frac{1}{2}\left[\exp\left\{-ar^2e^{2i\theta}(1+i\omega)\right\}+\exp\left\{-a r^2e^{2i\theta}(1-i\omega)\right\}\right] \\
&= \frac{1}{2} \big[\exp\left\{-a(\cos{2\theta}-\omega\sin{2\theta}+i\omega\cos{2\theta}+i\sin{2\theta})r^2\right\} \\
&\hspace{0.5cm}
+\exp\left\{-a(\cos{2\theta}+\omega\sin{2\theta}-i\omega\cos{2\theta}+i\sin{2\theta})r^2\right\}\big] \\
&\rightarrow \infty \ \ \ \text{for} \ \cos{2\theta}-\omega\sin{2\theta} < 0
\end{split}
\label{eq:CSCG}
\end{equation}
for positive values of the scaling angle $\theta$ and the parameter $\omega$.
To avoid this singularity, the complex-range Gaussian basis functions with a finite value of $\omega$ require a smaller scaling angle compared to the real-range Gaussian basis functions, in which the range of the scaling angle is $0\le \theta < \pi/2$.
The divergent behavior in the basis function might be serious when we calculate the long-range operator such as of the $E1$ transition.
In the present work, we take $\theta=12$ degrees and $\omega=1$ to avoid the divergence of the basis functions and to keep the numerical accuracy in the calculation.

\begin{figure}[tb]
\centering{\includegraphics[clip,width=8cm]{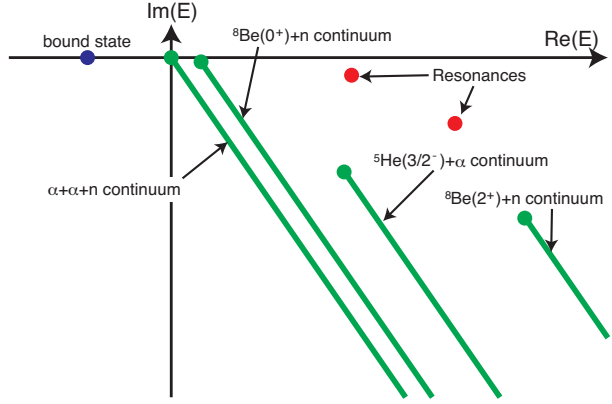}}
\caption{\label{fig:edis_sch}
(Color online) Schematic picture of energy eigenvalue distribution on the complex energy plane for the $\alpha$~+~$\alpha$~+~$n$ system.
}
\end{figure}
The energy eigenvalues $\{E^\theta_\nu\}$ are obtained on the complex energy plane, governed by the ABC theorem~\cite{Aguilar71,Balslev71}.
A schematic picture of the energy eigenvalue distribution is shown in Fig.~\ref{fig:edis_sch}.
In the CSM, the energies of bound states are given by real numbers and are invariant under the complex scaling.
On the other hand, resonances and continuum states are obtained as eigenstates with complex energy eigenvalues.
The resonances are obtained as isolated eigenstates on the complex energy plane, whose energies are given as $E^\theta_\nu = E^r_\nu-i\Gamma_\nu/2$.
The resonance energies $E^r_\nu$ and the decay widths $\Gamma_\nu$ are independent to the scaling angle $\theta$.
The continuum states are obtained on branch cuts rotated down by 2$\theta$ as shown in Fig.~\ref{fig:edis_sch}.
The branch cuts start from the different thresholds for two- and three-body continuum states in the case of the $\alpha$~+~$\alpha$~+~$n$ system as shown in Fig.~\ref{fig:edis_sch}.
This classification of the continuum states is useful in discussing the decay modes of the $^9$Be photodisintegration.

Using the energy eigenvalues and eigenstates of the complex-scaled Hamiltonian $\hat{H}^\theta$, we define the complex-scaled Green's function with outgoing boundary conditions, $\mathcal{G}^\theta(E; \boldsymbol{\xi}, \boldsymbol{\xi}')$, as
\begin{equation}
\mathcal{G}^\theta(E; \boldsymbol{\xi}, \boldsymbol{\xi}')
= \left\langle \boldsymbol{\xi} \left| \frac{1}{E-H^\theta} \right| \boldsymbol{\xi}' \right\rangle
= \sum_\nu\hspace{-0.46cm}\int \frac{\Psi^\nu(\theta) \tilde{\Psi}^\nu(\theta)}{E-E^\theta_\nu}.
\label{eq:CSGF}
\end{equation}
In the derivation of the right-hand side of Eq.~(\ref{eq:CSGF}), we use the extended completeness relation, whose detailed explanation is given in Ref.~\cite{Myo97}.
Using the complex-scaled Green's function, we calculate the photodisintegration cross section of $^9$Be into the $\alpha$~+~$\alpha$~+~$n$ scattering states.

We calculate the cross section of the photodisintegration of $^{9}\textrm{Be}(3/2^{-})~+~\gamma\rightarrow \alpha~+~\alpha~+~n$ in terms of the multipole response.
In the present calculation, we focus on the low-lying region of the photodisintegration cross section and take into account only the electromagnetic dipole responses.
The photodisintegration cross section $\sigma^\gamma$ is given by the sum of those by the $E1$ and $M1$ transition as
\begin{equation}
\sigma^\gamma(E_\gamma) = \sigma_{E1}(E_\gamma) + \sigma_{M1} (E_\gamma),
\end{equation}
where $E_{\gamma}$ is the incident photon energy.
The energy $E$ is related to $E_\gamma$ as $E=E_\gamma-E_\text{gs}$, where $E_\text{gs}$ is the binding energy of the $^9$Be ground state measured from the $\alpha$~+~$\alpha$~+~$n$ threshold.
The cross sections for the electromagnetic dipole transitions $\sigma_{EM1}$ is expressed as the following form;
\begin{equation}
\sigma_{EM1} (E_\gamma) =\frac{16\pi^3}{9}\cdot \left(\frac{E_{\gamma}}{\hbar c}\right) \cdot \frac{dB(EM1,E_{\gamma})}{dE_{\gamma}}.
\label{eq:phCS}
\end{equation}
Using the CSM and the complex-scaled Green's function in Eq.~(\ref{eq:CSGF}), the electromagnetic dipole transition strength is given as
\begin{equation}
\begin{split}
\frac{dB(EM1,E_\gamma)}{dE_\gamma} =& -\frac{1}{\pi} \cdot \frac{1}{2J_\text{gs}+1} \\
\times& \text{Im} \Bigg[\sum_\nu\hspace{-0.46cm}\int
\left\langle \tilde{\Psi}_\text{gs} || (\hat{O}^\theta)^\dagger(EM1) || \Psi^\nu(\theta) \right\rangle \\
&\times\frac{1}{E-E^\theta_\nu}
\left\langle \tilde{\Psi}^\nu(\theta) || \hat{O}^\theta(EM1) || \Psi_\text{gs} \right\rangle \Bigg],
\end{split}
\label{eq:E1}
\end{equation}
where $J_\text{gs}$ and $\Psi_\text{gs}$ represent the total spin and the wave function of the ground state, respectively, and $\hat{O}(EM1)$ is an electromagnetic dipole transition operator.

\section{Results\label{sec:r}}

\subsection{Ground-state properties}
The calculated ground-state properties of $^9$Be are listed in TABLE~\ref{ta:gspro}, in which $v_3$ and $\mu$ are taken as 1.10 MeV and 0.02 fm$^{-2}$, respectively, as mentioned in Sec.~\ref{sec:f}.
The binding energy and the charge radius are well reproduced by employing the three-body potential, while the calculated matter radius is slightly larger than the experimental value.

To see the ground-state structure of $^9$Be more in detail, we calculate the probability of each component of the $^8$Be($J^\pi$)~$\otimes$~$\nu l_j$ configurations.
The calculated probabilities are also listed in TABLE~\ref{ta:gspro}.
The present calculation shows that the valence neutron around $^8$Be occupies the $p_{3/2}$ orbit by 93.8~\%, while the excited $2^+$ component of $^8$Be has comparable amount to the $0^+$ one.
This large mixture of the $2^+$ component is understood using the $(\lambda,\mu) = (3,1)$ component in the SU(3) representation, which corresponds to the $\pi$-orbital of the valence neutron.
In the SU(3) representation, the (3,1) component gives 53~\% and 47~\% for the $^8$Be($0^+$)~$\otimes$~$\nu p$ and $^8$Be($2^+$)~$\otimes$~$\nu p$ components, respectively.
\begin{table}[tb]
\caption{\label{ta:gspro}
Ground-state properties in comparison with experiments.
The calculated binding energies ($E_\text{gs}$, unit in MeV), charge radii ($R_\text{ch}$, unit in fm), and matter ones ($R_\text{m}$, unit in fm).
The probabilities of the $^8$Be($J^\pi$) $\otimes$ $\nu l_j$ components in the ground state of $^9$Be are also shown.
}
\begin{ruledtabular}
\begin{tabular}{ccc}
& Present & Exp. \\
\hline
$E_\text{gs}$ & 1.57 & 1.5736\footnote{Reference \cite{Tilley04}} \\
$R_\text{ch}$ & 2.53 & 2.519$\pm$0.012\footnote{Reference \cite{Nortershauser09}} \\
$R_\text{m}$ & 2.42 & 2.38$\pm$0.01\footnote{Reference \cite{Tanihata88}} \\
\hline
$^8$Be($0^+$) $\otimes$ $\nu p_{3/2}$ & 47.06 \% & \\
$^8$Be($2^+$) $\otimes$ $\nu p_{3/2}$ & 46.77 \% & \\
$^8$Be($2^+$) $\otimes$ $\nu p_{1/2}$ & 2.31 \% & \\
$^8$Be($2^+$) $\otimes$ $\nu f_{7/2}$ & 1.21 \% & \\
$^8$Be($2^+$) $\otimes$ $\nu f_{5/2}$ & 1.20 \% & \\
$^8$Be($4^+$) $\otimes$ $\nu f_{7/2}$ & 1.07 \% & \\
$^8$Be($4^+$) $\otimes$ $\nu f_{5/2}$ & 0.25 \% & \\
$^8$Be($4^+$) $\otimes$ $\nu h_{11/2}$ & 0.04 \% & \\
$^8$Be($4^+$) $\otimes$ $\nu h_{9/2}$ & 0.04 \% & 
\end{tabular}
\end{ruledtabular}
\end{table}

\subsection{Photodisintegration cross section and resonances in low energy region}

We calculate the photodisintegration cross section of $^9$Be using Eq.~(\ref{eq:phCS}).
Before calculating the cross section, we determine the values of $v_3$ for each spin-parity state as shown in Table~\ref{ta:3bpot}.
It is noted that the strength for the $1/2^+$ state is slightly weakened from that in our previous work~\cite{Odsuren15}.
This comes from the inclusion of other spin-parity states and the improvement of the numerical accuracy in the low energy region by using the complex-range Gaussian basis functions.
\begin{table}[tb]
\caption{\label{ta:3bpot}
Strength of the three-body potential $v_3$ for each spin-parity state.
}
\begin{ruledtabular}
\begin{tabular}{ccccccc}
& $1/2^+$ & $3/2^+$ & $5/2^+$ & $1/2^-$ & $3/2^-$ & $5/2^-$ \\
\hline
$v_3$ (MeV) & -0.90 & -0.30 & -0.30 & 0.30 & 1.10 & 0.35 \\
\end{tabular}
\end{ruledtabular}
\end{table}

\begin{figure}[tb]
\centering{\includegraphics[clip,width=0.5\textwidth]{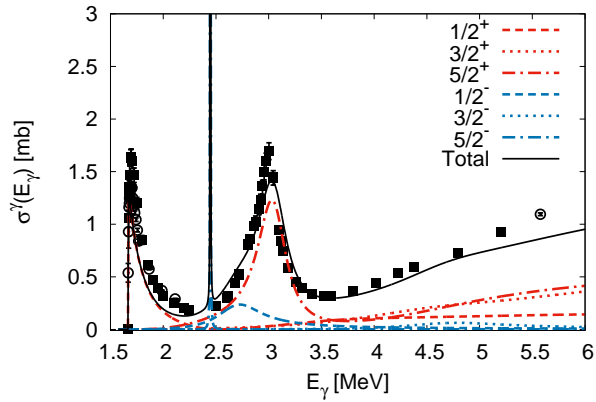}}
\caption{\label{fig:phCS}
(Color online) Calculated photodisintegration cross sections in comparison with the experimental data.
The solid squares and open circles represent the experimental data taken from Refs.~\cite{Arnold12} and \cite{Utsunomiya15}.
}
\end{figure}

In Fig.~\ref{fig:phCS}, the calculated photodisintegration cross section is shown in comparison with experimental data.
Our result well reproduces the observed cross section, not only the peak positions but also the magnitudes and widths of the peaks.
This agreement implies that our three-body model well describes the scattering states of $^9$Be in the low energy region.
The lowest peak just above the $^8$Be($0^+$)~+~$n$ threshold energy ($E_\gamma = 1.6654$ MeV) comes from the $E1$ transition into the $1/2^+$ state, and the calculated cross section shows that the strength below the $^8$Be($0^+$)~+~$n$ threshold is negligibly small.
The second and third peaks at $E_\gamma =$ 2.5 and 3.0 MeV in the cross section come from the transitions into the resonances of $5/2^-$ and $5/2^+$ states, respectively.
The $M1$ transition into the $1/2^-$ resonance has a sizable contribution to the cross section at around $E_\gamma =$ 2.7 MeV.
The transitions into the $3/2^\pm$ states play minor roles in the photodisintegration cross section below $E_\gamma =$ 6 MeV.

\begin{figure}[tb]
\centering{\includegraphics[clip,width=8.5cm]{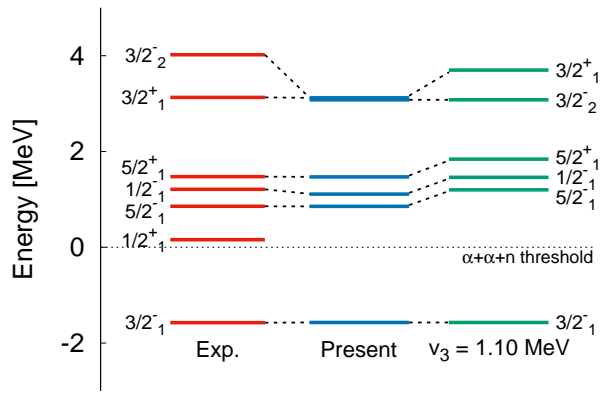}}
\caption{\label{fig:lev}
(Color online) Energy level diagram in the present calculation in comparison with the experimental data. The levels obtained with $v_3=1.10$ MeV for all the spin-parity states are also presented.
The experimental data are taken from Ref.~\cite{Tilley04}.}
\end{figure}

Using the CSM, we obtain the resonances corresponding to the isolated poles of the $S$-matrix as complex energy eigenvalues.
Our results of the resonance energies and decay widths for the excited states up to $E_\gamma = 6$ MeV are listed in TABLE~\ref{ta:res} in comparison with the experimental data.
The energy level diagram is also shown in Fig.~\ref{fig:lev}.
In Fig.~\ref{fig:lev}, we show the levels which are obtained using $v_3 = 1.10$ MeV commonly for all the spin-parity states for reference.
The calculated resonance energies and decay widths show good correspondences to the observed data, while the resonance energies of the excited states are shifted up by applying more repulsive three-body potential than those in Table.~\ref{ta:3bpot}. 
However, we cannot find any isolated resonance of the $1/2^+$ state in the present calculation, while the photodisintegration into the $1/2^+$ states has a significant peak above the $^8$Be($0^+$)~+~$n$ threshold.
These facts are consistent with our previous result~\cite{Odsuren15}.
It is noted that we confirm the $1/2^+$ state has the virtual-state character of the $^8$Be~+~$n$ system from a similar analysis to that in Ref.~\cite{Odsuren15}.
\begin{table}[tb]
\caption{\label{ta:res}
Resonance energies $E^r$ and decay widths $\Gamma$ for low-lying excited states (units in MeV).
The resonance energies are measured from the $\alpha$~+~$\alpha$~+~$n$ threshold.
The observed data are taken from Ref.~\cite{Tilley04}.
}
\begin{ruledtabular}
\begin{tabular}{ccccccc}
$J^\pi$ & Present ( $E^r$, $\Gamma$ ) & Exp. ( $E^r$, $\Gamma$ ) \\
\hline
$1/2^+_1$ & - & ( 0.158$\pm$0.002, 0.213$\pm$0.006 )\footnote{This value is taken from Ref.~\cite{Arnold12}} \\
$5/2^-_1$ & ( 0.854, $\sim$3$\times$$10^{-4}$ ) & ( 0.8558, 7.8$\times$$10^{-4}$ ) \\
$1/2^-_1$ & ( 1.11, 0.495 ) & ( 1.21, 1.01 ) \\
$5/2^+_1$ & ( 1.47, 0.323 ) & ( 1.475, 0.282 ) \\
$3/2^+_1$ & ( 3.12, 1.44 ) & ( 3.13, 0.743 ) \\
$3/2^-_2$ & ( 3.08, 1.18 ) & ( 4.02, 1.33 )
\end{tabular}
\end{ruledtabular}
\end{table}

\subsection{Low-lying dipole strength at $E_\gamma \sim 8$ MeV}

\begin{figure}[tb]
\centering{\includegraphics[clip,width=0.5\textwidth]{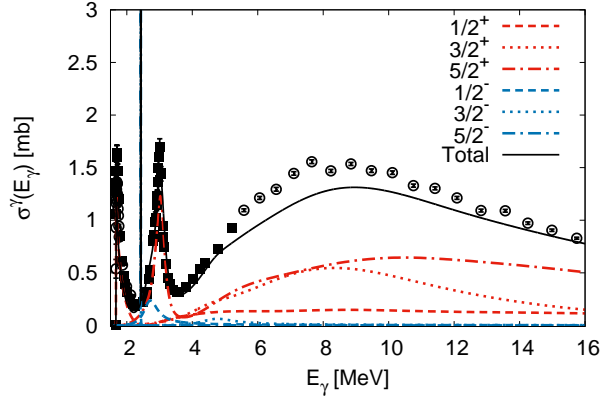}}
\caption{\label{fig:phCSL}
(Color online) Calculated photodisintegration cross section up to $E_\gamma =$ 16 MeV.
The solid squares and open circles represent experimental data taken from Refs.~\cite{Arnold12} and \cite{Utsunomiya15}, respectively.
}
\end{figure}

We investigate the photodisintegration cross section above $E_\gamma =$ 6 MeV and discuss the dipole strength of $^9$Be below the GDR.
In Fig.~\ref{fig:phCSL}, we show the calculated photodisintegration cross section up to $E_\gamma =$ 16 MeV in comparison with the experimental data.
From the result in Fig.~\ref{fig:phCSL}, we see that calculated cross section has a broad peak at $E_\gamma \sim$ 8 MeV as similar to the experimental data and the peak is dominated by the $E1$ transitions into $3/2^+$ and $5/2^+$.
In the present CSM calculation, we do not find any isolated resonances corresponding to the peak.
Therefore, the peak at $E_\gamma \sim 8$ MeV is understood to be described by the $E1$ transition from the ground state into the non-resonant continuum states of $3/2^+$ and $5/2^+$.
It is also seen that our calculation slightly underestimates the observed peak in the cross section.
This underestimation suggests a contribution from degrees of freedom beyond the relative motion of the $\alpha$~+~$\alpha$~+~$n$ system in the present model.
For example, the GDR is known to be described by coherent 1p-1h excitations, but the 1p-1h excitations from the $\alpha$ particle are not taken into account in our $\alpha$~+~$\alpha$~+~$n$ three-body model.

In the recent measurement, it is reported that the energy-integrated strength for the enhanced dipole strength newly measured for $4 \le E_\gamma \le 16$ MeV is estimated to be 11.3 mb MeV as a lower limit~\cite{Utsunomiya15}.
In Ref.~\cite{Utsunomiya15}, it is suggested that this energy-integrated strength exhausts 10~\% of the Thomas-Reiche-Kuhn (TRK) sum rule and almost all of the energy-weighted cluster dipole sum rule.
We calculated the corresponding energy-integrated strengths integrating the cross section $\sigma_{E1}$ for the electric dipole transition over the energy interval of $4 \le E_\gamma \le 16$ MeV.
We obtain the energy-integrated strength as 12.1 mb MeV for this energy interval, and the result is consistent with the experimental value (11.3 mb MeV).
On the other hand, we also obtain the energy-integrated strengths for $E_\gamma < 4$ MeV and $E_\gamma > 16$ MeV as 0.954 and 8.40 mb MeV, respectively.
It is seen that the energy-integrated strength for $4 \le E_\gamma \le 16$ MeV is 56.5~\% of the energy-weighted cluster dipole sum rule in the present calculation.

\begin{figure}[tb]
\centering{\includegraphics[clip,width=0.5\textwidth]{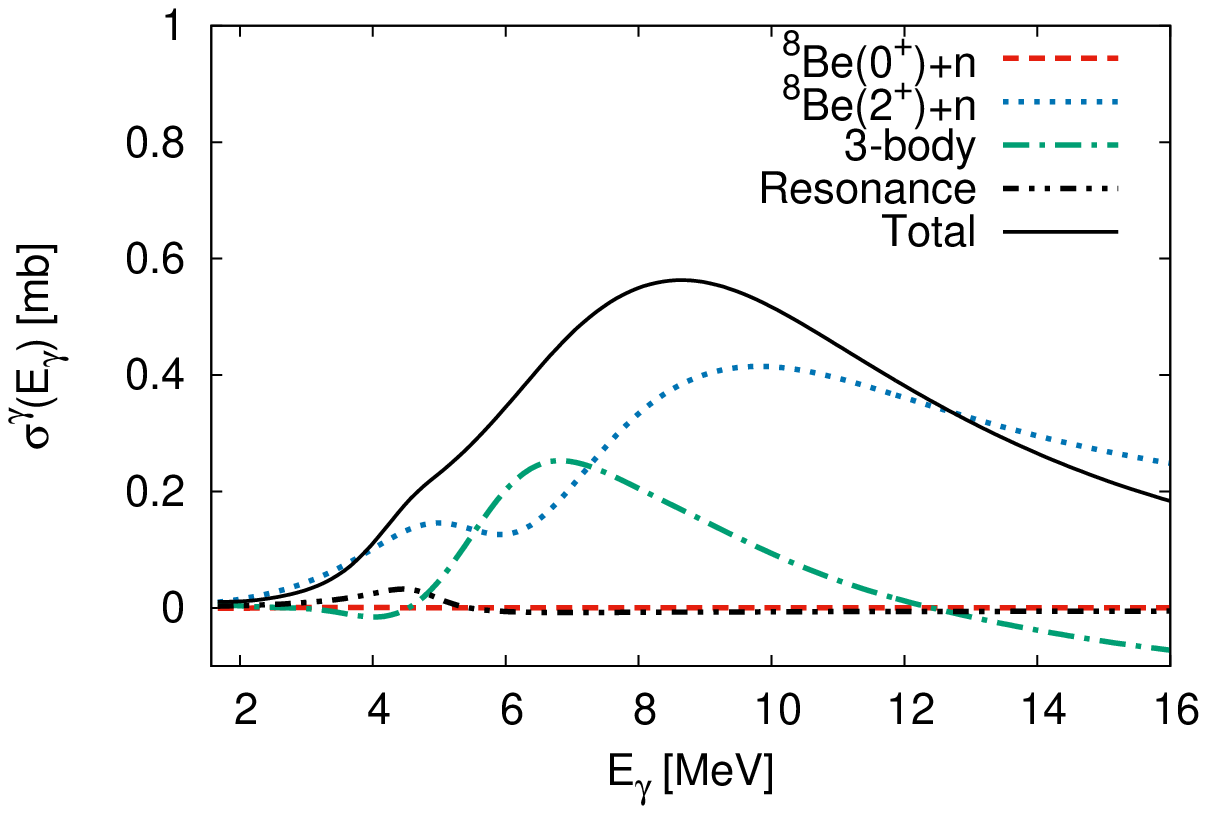}}
\caption{\label{fig:dec_3}
(Color online) Decomposed photodisintegration cross section for the $3/2^+$ states.
}
\centering{\includegraphics[clip,width=0.5\textwidth]{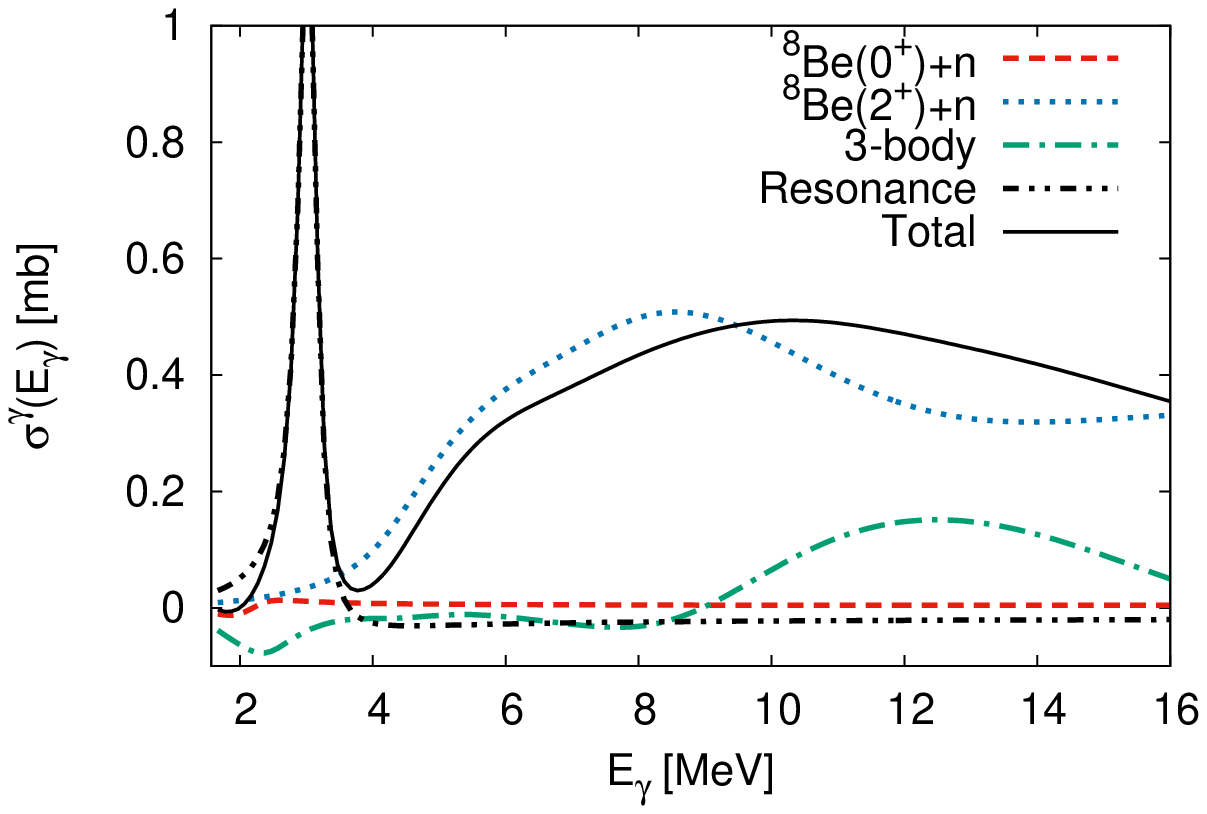}}
\caption{\label{fig:dec_5}
(Color online) Same as Fig.~\ref{fig:dec_3} but for the $5/2^+$ states.
}
\end{figure}

In the CSM, the energy eigenvalues of unbound states are classified into those of the resonance and several families of the two- and three-body non-resonant continuum states as shown in Fig.~\ref{fig:edis_sch}.
Combing the energy eigenvalues in the CSM with Eq.~(\ref{eq:E1}), we decomose the photodisintegration cross section into different families of the non-resonant continuum states.
In Figs~\ref{fig:dec_3} and \ref{fig:dec_5}, we show the decomposed photodisintegration cross sections for the $3/2^+$ and $5/2^+$ states, respectively.
We do not show the $^5$He~+~$n$ components in Figs.~\ref{fig:dec_3} and \ref{fig:dec_5}, since their contributions are much smaller than those from other components.
In both results, we confirmed that the cross sections at $E_\gamma \sim 8$ MeV are dominated by the contributions from the $^8$Be($2^+$)~+~$n$ non-resonant continuum states, while that from the $^8$Be($0^+$)~+~$n$ ones is negligible in the cross section.
Furthermore, we see that the three-body continuum states of $\alpha$~+~$\alpha$~+~$n$ have sizable contributions to the low-lying dipole strengths.

The decomposed photodisintegration cross sections show that the enhanced dipole strength at $E_\gamma \sim 8$  MeV is described by the single-neutron excitation from the $^8$Be($2^+$)~$\otimes$~$\nu p_{3/2}$ configuration in the ground state because the $E1$ transition operator cannot directly excite the relative motion between two $\alpha$'s from the $0^+$ to $2^+$ states.
The ground-state wave function has the large (3,1) component in the SU(3) representation, and the (3,1) component gives a mixture of the $^8$Be(0$^+$)~$\otimes$~$\nu p$ and $^8$Be(2$^+$)~$\otimes$~$\nu p$ configurations.
Our result shows that the ground-state structure of $^9$Be is essential to reproduce the observed dipole strength at $E_\gamma \sim 8$ MeV.

On the other hand, the contribution from the non-resonant continuum states of $^8$Be(0$^+$)~+~$n$ is negligibly small in the photodisintegration cross section.
This fact seems to be that the $^8$Be(0$^+$)~$\otimes$~$\nu p_{3/2}$ configuration is mainly excited into the resonances.
To confirm this, it would be important to investigate the structures of the excited resonances of $^9$Be and the further analysis will be performed in forthcoming paper.


\section{Summary\label{sec:s}}

In this work, we investigate the photodisintegration of $^9$Be in the energy region lower than $E_\gamma = 16$ MeV by using the $\alpha$~+~$\alpha$~+~$n$ three-body model and the complex scaling method (CSM).
We here employ the complex-range Gaussian basis functions, which enables us to describe the oscillating behaviors of the scattering states of the $\alpha$~+~$\alpha$~+~$n$ system and to calculate the transition strength accurately.
We calculate the photodisintegration cross section up to $E_\gamma = 16$ MeV and show good agreement with the observed data.
From the calculated cross section, we discuss the following two points:
One is the structures of the excited states of $^9$Be observed in the photodisintegration in the low energy region up to $E_\gamma = 6$ MeV.
The other is the origin of the enhanced dipole strength at $E_\gamma \sim 8$ MeV below the GDR.

In addition to the photodisintegration cross section, we also show the resonance energies and decay widths of the excited states obtained in the present calculation below $E_\gamma = 6$ MeV.
It is shown that the calculated resonance energies and decay widths are consistent with the observed energy levels while the the $1/2^+$ state is the exceptional case.
We confirm that the $1/2^+$ state has the virtual-state character of the $^8$Be~+~$n$ system from the similar analysis to that in our previous work.
The photodisintegration cross section up to $E_\gamma = 6$ MeV can be understood by the summation of transition strengths into each excited state of $^9$Be.

For the energy region of $6 \le E_\gamma \le 16$ MeV, our calculation shows a significant dipole strength at $E_\gamma \sim 8$ MeV and a good agreement with the recent observed data of the photodisintegration.
We find that this dipole strength is dominated by the transitions into the non-resonant continuum states of $3/2^+$ and $5/2^+$.
To understand the origin of the dipole strength, we decompose the photodisintegration cross section into each non-resonant continuum state by the use of the energy eigenvalue distribution in the CSM.
From the decomposed cross section, it is shown that the strength mainly comes from the transitions into the $^8$Be($2^+$)~+~$n$ non-resonant continuum states.
This fact can be understood by a dipole excitation of the single neutron from the $^8$Be($2^+$)~$\otimes$~$\nu p_{3/2}$ configuration in the $^9$Be ground state. 

In the present calculation, we do not investigate the structures of the excited resonances in detail.
It would be important to see the structures of the resonances to gain a comprehensive understanding of the photodisintegration of $^9$Be, and the detailed analysis of the resonances will be performed in the forthcoming paper.

\section*{Acknowledgements}
The authors are grateful to Prof. H.~Utsunomiya for fruitful discussion and providing us with the experimental data for the photodisintegration cross section of $^9$Be.
This work was supported by JSPS KAKENHI Grant Numbers 25400241 and 20423212.
One of the authors (K.~Kat\=o) also thanks for the support by the International Collaboration with Al-Farabi Kazakh National University (Grant No: 3106/GF4 and 1550/GF3).

\end{document}